\documentclass[11pt]{article}
\usepackage{amsfonts,amsmath,amssymb,cite}
\usepackage{graphicx}
\begin{document}
\title{Dirichlet-to-Neumann and Neumann-to-Dirichlet methods 
for bound states of the Helmholtz equation}
\author{Sebastian Bielski\\
Atomic Physics Division,
Department of Atomic Physics and Luminescence, \\
Faculty of Applied Physics and Mathematics, Gda{\'n}sk University of Technology,\\ Narutowicza 11/12, PL 80--233 Gda{\'n}sk, Poland\\
Email:bolo@mif.pg.gda.pl}
\date{\today}
\maketitle

\begin{abstract}
Two methods for computing bound states of the Helmholtz equation in a  finite domain are presented. 
The methods are formulated in terms of the Dirichlet-to-Neumann (DtN) and Neumann-to-Dirichlet (NtD) surface integral operators. They are adapted from the DtN and NtD methods for bound states of the Schr\"odinger equation in $\mathbb{R}^{3}$. 
A variational principle that enables the usage of the operators is constructed. 
The variational principle allows the use of discontinuous (in values or derivatives) trial functions. A numerical example presenting the usefulness of the DtN and NtD methods is given.
\end{abstract}
\maketitle

\section{Introduction}
In this work we are interested in finding solutions to the problem consisting of the Helmholtz equation defined in a two- (or more) 
dimensional finite volume $\Gamma$:
\begin{equation}
\Delta \Psi(\mathbf{r})+k^{2}\Psi(\mathbf{r})=0, \qquad 
\mathbf{r} \in \Gamma
\label{eq1}
\end{equation}
and the homogeneous Dirichlet condition on the boundary $\partial \Gamma$:
\begin{equation}
\Psi(\mathbf{r})=0, \qquad \mathbf{r} \in \partial \Gamma.
\label{eq2}
\end{equation}
The set (\ref{eq1})--(\ref{eq2}) is an eigenproblem in which the values $\left\{-k^{2}\right\}$ are the eigenvalues and $\left\{\Psi\right\}$ are the corresponding eigenfunctions. The eigenproblem (\ref{eq1})--(\ref{eq2}) appears in many different areas of physics. It describes, for example, the behaviour of a particle confined in an infinitely deep potential (in this case $k^{2}$ is proportional to the energy of the particle while $\left|\Psi\right|^{2}$ is the probability density) or vibrations of a homogeneous membrane ($k$ is proportional to the vibration frequency, $\Psi$ is the amplitude), it is useful in studying the propagation of electromagnetic waves in waveguides, etc. So, although the problem of finding eigenvalues and eigenfunctions of the Laplace operator has been known for many decades, it remains very important in many fields. 

The standard analytical approach to problems like (\ref{eq1})--(\ref{eq2}) is the method of separation of variables. The first step in the method is to choose an appropriate coordinate system. The choice depends on the shape of $\partial \Gamma$. In practice, only in some cases it is possible to find a system fitted to the geometry of a problem and to obtain the exact solutions using the separation of variables technique. In general, the shape of the boundary of $\Gamma$ may be arbitrary and no useful coordinate system may be found, so other methods may need to be used. There are many different attempts. 
Amore \cite{Amor08} has applied a collocation method using so-called little sinc functions for problems defined in two-dimensional domains of arbitrary shape. Chakraborty et al. \cite{Chak09} have presented an analytical perturbative method. In the two mentioned works brief surveys of other methods may be found. Recently, Steinbach et al. \cite{Stei10} have formulated a boundary element domain decomposition method that enables to transform the original problem to a new one defined on the boundaries separating the subdomains. 

The goal of this work is to present two methods that are applicable to the eigenproblem (\ref{eq1})--(\ref{eq2}) in case the domain is such that it can be naturally divided into two non-overlapping subdomains. The methods consist in the application of a variational principle allowing the use of trial functions that may experience jumps in values or derivatives when passing from one subdomain to other. The Dirichlet-to-Neumann (DtN) integral operator or the Neumann-to-Dirichlet (NtD) integral operator is used, both are defined on the interface separating the subdomains. Each of the methods allows to replace the initial problem (\ref{eq1})--(\ref{eq2}) with a new problem defined in one of the subdomains and on the interface. The methods are related to the DtN and the NtD embedding methods for the bound states of the Schr\"odinger equation (defined in $\mathbb{R}^{3}$) and their relativistic counterparts \cite{Szmy04, Biel06}. The DtN method for the Schr\"odinger equation is a close relative of the embedding method proposed by Inglesfield \cite{Ingl81}. In the Inglesfield's method the Green function formalism is used while in the DtN (and NtD) method an operator approach analogous to that employed in the R-matrix theory \cite{Szmy97, Szmy98} is applied.

The structure of the paper is as follows. In section II a systematic construction of a variational principle (allowing the use of discontinuous trial functions) for bound states of the Helmholtz equation is presented. In section III the DtN and NtD operators are defined. Sections IV and V are devoted to the formalism of the DtN and NtD methods for bound states of the Helmholtz equation. In section VI a numerical example is provided.

\section{Variational principle allowing the use of discontinuous trial functions}
\begin{figure}[h]
\begin{center}
\includegraphics[width=5cm]{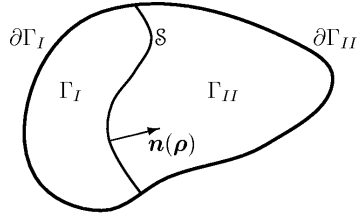}
\caption{\label{fig1}Partitioning of the domain $\Gamma$ into two subdomains $\Gamma_{I}$ and $\Gamma_{II}$, separated by the interface $\mathcal{S}$; $\mathbf{n}(\boldsymbol{\rho})$ is the unit vector normal to the interface $\mathcal{S}$ at the point $\boldsymbol{\rho}$.}
\end{center}
\end{figure}

Let $\Gamma$ be a two- (or more) dimensional finite domain of such a shape that it may be in a natural way divided into two subdomains, $\Gamma_{I}$ and $\Gamma_{II}$, separated by a smooth curve (or surface) denoted by $\mathcal{S}$, as shown in 
figure~\ref{fig1}. Thus, the boundary of $\Gamma_{I}$ consists of $\partial \Gamma_{I}$ and $\mathcal{S}$ while the boundary of $\Gamma_{II}$ is composed of $\partial \Gamma_{II}$ and $\mathcal{S}$. 
A position vector lying on $\mathcal{S}$ will be denoted by $\boldsymbol{\rho}$ and $\mathbf{n}(\boldsymbol{\rho})$ will be the unit vector normal to $\mathcal{S}$ at the point $\boldsymbol{\rho}$ (we assume that $\mathbf{n}(\boldsymbol{\rho})$ is always pointed outward from $\Gamma_{I}$). Denoting
\begin{equation}
\Psi_{i}(\mathbf{r}) 
= \Psi(\mathbf{r}) 
\qquad (\mathbf{r}\in\Gamma_{i}\,; i=I,\textit{II}),
\label{eq3}
\end{equation}
we may rewrite the initial problem (\ref{eq1})--(\ref{eq2}) as
\begin{equation}
\Delta \Psi_{I}(\mathbf{r})+k^{2}\Psi_{I}(\mathbf{r})=0, \qquad 
\mathbf{r} \in \Gamma_{I},
\label{eq4}
\end{equation}
\begin{equation}
\Psi_{I}(\mathbf{r})=0, \qquad \mathbf{r} \in \partial \Gamma_{I},
\label{eq5}
\end{equation}
\begin{equation}
\Delta \Psi_{II}(\mathbf{r})+k^{2}\Psi_{II}(\mathbf{r})=0, \qquad 
\mathbf{r} \in \Gamma_{II},
\label{eq6}
\end{equation}
\begin{equation}
\Psi_{II}(\mathbf{r})=0, \qquad \mathbf{r} \in \partial \Gamma_{II}.
\label{eq7}
\end{equation}
The function $\Psi(\mathbf{r})$ and its gradient must be continuous in the whole domain $\Gamma$, so 
it is obvious that the functions $\Psi_{I}(\mathbf{r})$ and $\Psi_{II}(\mathbf{r})$ obey the equations:
\begin{equation}
\Psi_{I}(\boldsymbol{\rho})-\Psi_{II}(\boldsymbol{\rho})=0,
\label{eq8}
\end{equation}
\begin{equation}
\nabla_{\perp}\Psi_{I}(\boldsymbol{\rho})
-\nabla_{\perp}\Psi_{II}(\boldsymbol{\rho})=0,
\label{eq9}
\end{equation}
where
\begin{equation}
\nabla_{\perp}\Psi_{i}(\boldsymbol{\rho})
=\mathbf{n}(\boldsymbol{\rho})\cdot\boldsymbol{\nabla}
\Psi_{i}(\mathbf{r})\Big|_{\mathbf{r}=\boldsymbol{\rho}}
\label{eq10}
\end{equation}
is the normal derivative of $\Psi_{i}$ at $\boldsymbol{\rho}$.

We want to determine the values of $\left\{k^{2}\right\}$ and the corresponding functions $\left\{\Psi(\mathbf{r})\right\}$. 
Basing on equations (\ref{eq4}), (\ref{eq6}), (\ref{eq8}) and (\ref{eq9}) and using a method proposed by Gerjuoy et al. \cite{Gerj83} we define a functional that provides some estimate of one of the sought values $\left\{k^{2}\right\}$:
\begin{eqnarray}
&& \mathcal{F}[\overline{k},\overline{\Psi}_{I},\overline{\Psi}_{II};
\overline{\Lambda}_{I},\overline{\Lambda}_{II},\overline{\lambda},
\overline{\chi}]=\overline{k}\,^{2}+\big<\overline{\Lambda}_{I}
\big|[\Delta+\overline{k}\,^{2}]
\overline{\Psi}_{I}\big>_{I}
+\big<\overline{\Lambda}_{II}\big|[\Delta+\overline{k}\,^{2}]
\overline{\Psi}_{II}\big>_{II}
\nonumber \\
&&  \qquad
+\big(\overline{\lambda}\big|\overline{\Psi}_{I}
-\overline{\Psi}_{II}\big)+\big(\overline{\chi}\big|
\nabla_{\perp}\overline{\Psi}_{I}-\nabla_{\perp}
\overline{\Psi}_{II}\big).
\label{eq11}
\end{eqnarray}
The scalar products are defined as follows:
\begin{equation}
\big<\Phi\big|\Phi'\big>_{i}
=\int_{\Gamma_{i}}\mathrm{d}\mathbf{r}\:
\Phi^{*}(\mathbf{r})\Phi'(\mathbf{r}),
\label{eq12}
\end{equation}
\begin{equation}
\big(\Phi\big|\Phi'\big)
=\int_{\mathcal{S}}\mathrm{d}\boldsymbol{\rho}\:
\Phi^{*}(\boldsymbol{\rho})\Phi'(\boldsymbol{\rho})
\label{eq13}
\end{equation}
($\mathrm{d}\mathbf{r}$ is an infinitesimal volume element around the point $\mathbf{r}$, $\mathrm{d}\boldsymbol{\rho}$ is an infinitesimal scalar element of the interface $\mathcal{S}$ around the point $\boldsymbol{\rho}$, and $*$ denotes the complex conjugation). The value $\overline{k}$, the function $\overline{\Psi}_{I}$ (vanishing on 
$\partial \Gamma_{I}$) and  the function 
$\overline{\Psi}_{II}$ (vanishing on $\partial \Gamma_{II}$) 
are some trial estimates of the exact quantities $k$, $\Psi_{I}$ and $\Psi_{II}$. The functions $\overline{\Lambda}_{I}$ (defined in $\Gamma_{I}$), 
$\overline{\Lambda}_{II}$ (defined in $\Gamma_{II}$), $\overline{\lambda}$ and $\overline{\chi}$ (both defined on $\mathcal{S}$) play role of the Lagrange functions including equations (\ref{eq4}), (\ref{eq6}), (\ref{eq8}) and (\ref{eq9}) in the functional. 
The first variation of the functional (\ref{eq11}) 
with respect to arbitrary variations of $\overline{k}$, $\overline{\Psi}_{I}$, $\overline{\Psi}_{II}$ about $k$, $\Psi_{I}$, $\Psi_{II}$ (supposing that the variations $\delta\Psi_{I}$ and $\delta\Psi_{II}$ vanish on $\partial \Gamma_{I}$ and $\partial \Gamma_{II}$, respectively) and $\overline{\Lambda}_{I}$, $\overline{\Lambda}_{II}$, $\overline{\lambda}$, 
$\overline{\chi}$ about some arbitrarily chosen $\Lambda_{I}$, $\Lambda_{II}$, $\lambda$, $\chi$ may be written as
\begin{eqnarray}
&& \delta \mathcal{F}[k,\Psi_{I},\Psi_{II};
\Lambda_{I},\Lambda_{II},\lambda,\chi]
= 2k \delta k \left[ 1+\big<\Lambda_{I}\big|\Psi_{I}\big>_{I}
+\big<\Lambda_{II}\big|\Psi_{II}\big>_{II} \right] 
\nonumber \\
&& \qquad
+\big<[\Delta+k^{2}]\Lambda_{I}\big|\delta\Psi_{I}\big>_{I} 
+\big<[\Delta+k^{2}]\Lambda_{II}\big|\delta\Psi_{II}\big>_{II} 
\nonumber \\
&& \qquad
+\big(\lambda - \nabla_{\perp}\Lambda_{I} 
\big|\delta\Psi_{I}\big)
-\big(\lambda - \nabla_{\perp}\Lambda_{II} 
\big|\delta\Psi_{II}\big)
\nonumber \\
&& \qquad
+\big(\chi+\Lambda_{I}\big|\nabla_{\perp}\delta\Psi_{I}\big)
-\big(\chi+\Lambda_{II}\big|\nabla_{\perp}\delta\Psi_{II}\big)
\nonumber \\
&& \qquad
+\big(\Lambda_{I}\big|\nabla_{\perp}\delta\Psi_{I}\big)_{I}
+\big(\Lambda_{II}\big|\nabla_{\perp}\delta\Psi_{II}\big)_{II},
\label{eq14}
\end{eqnarray}
where
\begin{equation}
\label{eq15}
\big(\Phi\big|\Phi'\big)_{i}
=\int_{\partial \Gamma_{i}}\mathrm{d}\mathbf{r}\:
\Phi^{*}(\mathbf{r})\Phi'(\mathbf{r}).
\end{equation}
In the above scalar product, $\mathrm{d}\mathbf{r}$ is an infinitesimal scalar element of $\partial \Gamma_{i}$ around the point $\mathbf{r}$ (cf. the definitions (\ref{eq12}) and (\ref{eq13})).
We seek such functions $\Lambda_{I}$, $\Lambda_{II}$, $\lambda$ and $\chi$ 
for which the functional is stationary, i.e.~its first variation is equal to zero. So the functions $\Lambda_{I}$, $\Lambda_{II}$, $\lambda$ and $\chi$ fulfil the equations:
\begin{equation}
1+\big<\Lambda_{I}\big|\Psi_{I}\big>_{I}
+\big<\Lambda_{II}\big|\Psi_{II}\big>_{II}=0,
\label{eq16}
\end{equation}
\begin{equation}
[\Delta+k^{2}]\Lambda_{i}(\mathbf{r})=0, \qquad 
\mathbf{r} \in \Gamma_{i},
\label{eq17}
\end{equation}
\begin{equation}
\Lambda_{i}(\mathbf{r})=0, \qquad 
\mathbf{r} \in \partial \Gamma_{i},
\label{eq18}
\end{equation}
\begin{equation}
\lambda(\boldsymbol{\rho}) - \nabla_{\perp}\Lambda_{i}(\boldsymbol{\rho})=0,
\label{eq19}
\end{equation}
\begin{equation}
\chi(\boldsymbol{\rho})+\Lambda_{i}(\boldsymbol{\rho})=0,
\label{eq20}
\end{equation}
where $i=I,\textit{II}$. From equations (\ref{eq19}) and (\ref{eq20}) we obtain
\begin{equation}
\Lambda_{I}(\boldsymbol{\rho})-\Lambda_{II}(\boldsymbol{\rho})=0,
\label{eq21}
\end{equation}
\begin{equation}
\nabla_{\perp}\Lambda_{I}(\boldsymbol{\rho})-
\nabla_{\perp}\Lambda_{II}(\boldsymbol{\rho})=0.
\label{eq22}
\end{equation}
Comparying equations (\ref{eq17}), (\ref{eq18}), (\ref{eq21}) and (\ref{eq22})
with equations (\ref{eq4})--(\ref{eq9}) we find that $\Lambda_{I}(\mathbf{r})$ and $\Lambda_{II}(\mathbf{r})$ obey the same differential equations and the same boundary conditions as $\Psi_{I}(\mathbf{r})$ and $\Psi_{II}(\mathbf{r})$. This means that the functions $\Lambda_{I}$ and $\Lambda_{II}$ are proportional to $\Psi_{I}$ and $\Psi_{II}$:
\begin{equation}
\Lambda_{i}(\mathbf{r})=\eta \Psi_{i}(\mathbf{r}) \qquad (i=I,\textit{II}).
\label{eq23}
\end{equation}
The value of $\eta$ may be found using the formulas (\ref{eq23}) in equation (\ref{eq16}):
\begin{equation}
\eta=-\frac{1}{\big<\Psi_{I}\big|\Psi_{I}\big>_{I}
+\big<\Psi_{II}\big|\Psi_{II}\big>_{II}}.
\label{eq24}
\end{equation}
According to equations (\ref{eq19}), (\ref{eq20}) and (\ref{eq23}), we may write
\begin{equation}
\lambda(\boldsymbol{\rho})=
\eta[a\nabla_{\perp}\Psi_{I}(\boldsymbol{\rho})
+(1-a)\nabla_{\perp}\Psi_{II}(\boldsymbol{\rho})],
\label{eq25}
\end{equation}
\begin{equation}
\chi(\boldsymbol{\rho})=-
\eta[b\Psi_{I}(\boldsymbol{\rho})+(1-b)\Psi_{II}(\boldsymbol{\rho})],
\label{eq26}
\end{equation}
where $a$ and $b$ are arbitrary complex constants.

Now, let us assume, that the trial Lagrange functions $\overline{\Lambda}_{I}$,$\overline{\Lambda}_{II}$,$\overline{\lambda}$, $\overline{\chi}$, appearing in the functional (\ref{eq11}), are related to the estimates  
$\overline{\Psi}_{I}$ and $\overline{\Psi}_{II}$ in the same way that the functions $\Lambda_{I}$, $\Lambda_{II}$, $\lambda$, $\chi$ are related to 
the exact functions $\Psi_{I}$ and $\Psi_{II}$.
Using the formulas obtained from equations (\ref{eq23})--(\ref{eq26}) by replacing the functions $\Psi_{I}$, $\Psi_{II}$, $\Lambda_{I}$, $\Lambda_{II}$, $\lambda$ and $\chi$ with the trial functions $\overline{\Psi}_{I}$, $\overline{\Psi}_{II}$, $\overline{\Lambda}_{I}$, $\overline{\Lambda}_{II}$, $\overline{\lambda}$ and $\overline{\chi}$, transforms the functional (\ref{eq11}) to
\begin{eqnarray}
&& \mathcal{F}[\overline{\Psi}_{I},\overline{\Psi}_{II}]
=-\frac{\big<\overline{\Psi}_{I}\big|
\Delta\overline{\Psi}_{I}\big>_{I}
+\big<\overline{\Psi}_{II}\big|
\Delta\overline{\Psi}_{II}\big>_{II}}
{\big<\overline{\Psi}_{I}\big|\overline{\Psi}_{I}\big>_{I}
+\big<\overline{\Psi}_{II}\big|
\overline{\Psi}_{II}\big>_{II}} 
-\frac{\big(a\nabla_{\perp}\overline{\Psi}_{I}
+[1-a]\nabla_{\perp}\overline{\Psi}_{II}\big|
\overline{\Psi}_{I}-\overline{\Psi}_{II}\big)}
{\big<\overline{\Psi}_{I}\big|\overline{\Psi}_{I}\big>_{I}
+\big<\overline{\Psi}_{II}\big|
\overline{\Psi}_{II}\big>_{II}}
\nonumber \\
&& \qquad
+\frac{\big(b\overline{\Psi}_{I}+[1-b]\overline{\Psi}_{II}\big|
\nabla_{\perp}\overline{\Psi}_{I}
-\nabla_{\perp}\overline{\Psi}_{II}\big)}
{\big<\overline{\Psi}_{I}\big|\overline{\Psi}_{I}\big>_{I}
+\big<\overline{\Psi}_{II}\big|
\overline{\Psi}_{II}\big>_{II}}.
\label{eq27}
\end{eqnarray}
Our functional is supposed to estimate of a real quantity, so it should possess the property
\begin{equation}
\mathcal{F}^{*}[\overline{\Psi}_{I},\overline{\Psi}_{II}]
=\mathcal{F}[\overline{\Psi}_{I},\overline{\Psi}_{II}].
\label{eq28}
\end{equation}
After some rearrangements we find that equation (\ref{eq28}) is obeyed if
\begin{equation}
b=1-a^{*}
\label{eq29}
\end{equation}
and the final form of the functional is
\begin{eqnarray}
&& \mathcal{F}[\overline{\Psi}_{I},\overline{\Psi}_{II}]
=-\frac{\big<\overline{\Psi}_{I}\big|
\Delta\overline{\Psi}_{I}\big>_{I}
+\big<\overline{\Psi}_{II}\big|
\Delta\overline{\Psi}_{II}\big>_{II}}
{\big<\overline{\Psi}_{I}\big|\overline{\Psi}_{I}\big>_{I}
+\big<\overline{\Psi}_{II}\big|
\overline{\Psi}_{II}\big>_{II}} 
-\frac{\big(a\nabla_{\perp}\overline{\Psi}_{I}
+[1-a]\nabla_{\perp}\overline{\Psi}_{II}\big|
\overline{\Psi}_{I}-\overline{\Psi}_{II}\big)}
{\big<\overline{\Psi}_{I}\big|\overline{\Psi}_{I}\big>_{I}
+\big<\overline{\Psi}_{II}\big|
\overline{\Psi}_{II}\big>_{II}}
\nonumber \\
&& \qquad
+\frac{\big([1-a^{*}]\overline{\Psi}_{I}+a^{*}\overline{\Psi}_{II}\big|\nabla_{\perp}\overline{\Psi}_{I}
-\nabla_{\perp}\overline{\Psi}_{II}\big)}
{\big<\overline{\Psi}_{I}\big|\overline{\Psi}_{I}\big>_{I}
+\big<\overline{\Psi}_{II}\big|
\overline{\Psi}_{II}\big>_{II}}.
\label{eq30}
\end{eqnarray}
It is easy to verify that the exact functions are the stationary points of the functional (\ref{eq30}):
\begin{equation}
\delta\mathcal{F}[\Psi_{I},\Psi_{II}]=0,
\label{eq31}
\end{equation}
and the corresponding stationary values are equal to $k^{2}$:
\begin{equation}
\mathcal{F}[\Psi_{I},\Psi_{II}]=k^{2}.
\label{eq32}
\end{equation}
The initial problem (\ref{eq1})--(\ref{eq2}) is then equivalent to the variational principle (\ref{eq30})--(\ref{eq32}). The important thing is that the functional (\ref{eq30}) allows to use trial functions $\overline{\Psi}_{I}$ and $\overline{\Psi}_{II}$ that, together with their gradients, are continuous in their domains, but do not have to match at $\mathcal{S}$, so at least one of the following cases may occur:
\begin{equation}
\overline{\Psi}_{I}(\boldsymbol{\rho})\neq\overline{\Psi}_{II}(\boldsymbol{\rho}), \quad \nabla_{\perp}\overline{\Psi}_{I}(\boldsymbol{\rho})\neq
\nabla_{\perp}\overline{\Psi}_{II}(\boldsymbol{\rho}).
\label{eq33}
\end{equation}
It is worth noting that such functionals are rather rarely applied. 
More details about the construction of similar functionals may be found in the paper of Szmytkowski et al. \cite{Szmy04a}, where variational principles for bound states of the Schr\"odinger and the Dirac equations have been  presented.

\section{The DtN and NtD operators}
Let us assume that the subdomain $\Gamma_{II}$ is such that we are able to find analytically the functions $\psi(\kappa,\mathbf{r})$ obeying the Helmholtz equation at some fixed real value of the parameter $\kappa$ (which need not be in the spectrum of the eigenproblem (\ref{eq1})--(\ref{eq2})):
\begin{equation}
\Delta \psi(\kappa,\mathbf{r})+\kappa^{2}\psi(\kappa,\mathbf{r})=0, \qquad 
\mathbf{r} \in \Gamma_{II}
\label{op1}
\end{equation}
and the boundary condition
\begin{equation}
\psi(\kappa,\mathbf{r})=0, \qquad \mathbf{r} \in \partial \Gamma_{II}.
\label{op2}
\end{equation}
Now, let us define two mutually reciprocal integral operators $\hat{\mathcal{B}}(\kappa)$ and $\hat{\mathcal{R}}(\kappa)$ such that for every $\psi(\kappa,\mathbf{r})$ 
at the interface $\mathcal{S}$ it holds that
\begin{equation}
\nabla_{\perp}\psi(\kappa,\boldsymbol{\rho})
=\hat{\mathcal{B}}(\kappa)\psi(\kappa,\boldsymbol{\rho}),
\label{op3}
\end{equation}
\begin{equation}
\hat{\mathcal{R}}(\kappa)
\nabla_{\perp}\psi(\kappa,\boldsymbol{\rho})
=\psi(\kappa,\boldsymbol{\rho})
\label{op4}
\end{equation}
(note that the operators $\nabla_{\perp}$ and $\hat{\mathcal{B}}(\kappa)$ are not identical, equation (\ref{op3}) is valid only for the functions $\psi(\kappa,\mathbf{r})$). 
The operator $\hat{\mathcal{B}}(\kappa)$ transforms the Dirichlet datum 
$\psi(\kappa,\boldsymbol{\rho})$ into the Neumann datum $\nabla_{\perp}\psi(\kappa,\boldsymbol{\rho})$ so it is called the Dirichlet-to-Neumann (DtN) operator. In analogy, the operator $\hat{\mathcal{R}}(\kappa)$ is called the Neumann-to-Dirichlet (NtD) operator. Using integral kernels $\mathcal{B}(\kappa,\boldsymbol{\rho},\boldsymbol{\rho}')$ and $\mathcal{R}(\kappa,\boldsymbol{\rho},\boldsymbol{\rho}')$ of the operators, we may rewrite equations (\ref{op3}) and (\ref{op4}) in the following forms:
\begin{equation}
\nabla_{\perp}\psi(\kappa,\boldsymbol{\rho})
=\int_{\mathcal{S}}\mathrm{d}\boldsymbol{\rho}'\:
\mathcal{B}(\kappa,\boldsymbol{\rho},\boldsymbol{\rho}')
\psi(\kappa,\boldsymbol{\rho}'),
\label{op5}
\end{equation}
\begin{equation}
\psi(\kappa,\boldsymbol{\rho})
=\int_{\mathcal{S}}\mathrm{d}\boldsymbol{\rho}'\:
\mathcal{R}(\kappa,\boldsymbol{\rho},\boldsymbol{\rho}')
\nabla_{\perp}\psi(\kappa,\boldsymbol{\rho}').
\label{op6}
\end{equation}
Now, let us analyze the eigensystem
\begin{equation}
\Delta \psi_{n}(\kappa,\mathbf{r})+\kappa^{2}\psi_{n}(\kappa,\mathbf{r})=0, \qquad \mathbf{r} \in \Gamma_{II},
\label{op7}
\end{equation}
\begin{equation}
\psi_{n}(\kappa,\mathbf{r})=0, \qquad \mathbf{r} \in \partial \Gamma_{II},
\label{op8}
\end{equation}
\begin{equation}
\nabla_{\perp}\psi_{n}(\kappa,\boldsymbol{\rho})
=b_{n}(\kappa)\psi_{n}(\kappa,\boldsymbol{\rho}).
\label{op9}
\end{equation}
In the above system $b_{n}(\kappa)$ is an eigenvalue, $\psi_{n}(\kappa,\mathbf{r})$ is an eigenfunction and $\kappa$ is some fixed real parameter. 
The eigensystem (\ref{op7})--(\ref{op9}) is non-standard, because the eigenvalue $b_{n}(\kappa)$ appears not in the differential equation but in the boundary condition. Eigenproblems of such type are known as the Steklov eigenproblems \cite{Auch04}.

Using the Green's theorem (and the condition (\ref{op8})) for two arbitrary eigenfunctions 
$\psi_{n}(\kappa,\mathbf{r})$ and $\psi_{n'}(\kappa,\mathbf{r})$ we obtain
\begin{equation}
\big<\psi_{n}\big|\Delta\psi_{n'}\big>_{II}
-\big<\Delta\psi_{n}\big|\psi_{n'}\big>_{II}
=\big(\nabla_{\perp}\psi_{n}\big|\psi_{n'}\big)
-\big(\psi_{n}\big|\nabla_{\perp}\psi_{n'}\big).
\label{op10}
\end{equation}
In virtue of equation (\ref{op7}) the left-hand side of equation (\ref{op10}) vanishes. Applying equation (\ref{op9}) leads us to
\begin{equation}
[b_{n}^{*}(\kappa)-b_{n'}(\kappa)]
\big(\psi_{n}\big|\psi_{n'}\big)=0.
\label{op11}
\end{equation}
There are two conclusions we may draw from equation (\ref{op11}). First, if we take $n'$ equal to $n$, we see that the eigenvalues are real
\begin{equation}
b_{n}(\kappa)=b_{n}^{*}(\kappa).
\label{op12}
\end{equation}
Second, if the eigenfunctions $\psi_{n}(\kappa,\mathbf{r})$ and $\psi_{n'}(\kappa,\mathbf{r})$ belong to different eigenvalues, they are orthogonal with respect to the surface scalar product (\ref{eq13}):
\begin{equation}
\big(\psi_{n}\big|\psi_{n'}\big)=0
\qquad [b_{n}(\kappa)\neq b_{n'}(\kappa)].
\label{op13}
\end{equation}
Now, let us assume that all the eigenfunctions of (\ref{op7})--(\ref{op9}) are orthonormal on $\mathcal{S}$:
\begin{equation}
\big(\psi_{n}\big|\psi_{n'}\big)=\delta_{nn'},
\label{op14}
\end{equation}
and that the surface functions $\left\{\psi_{n}(\kappa,\boldsymbol{\rho})\right\}$ form a complete set in the space of single-valued square-integrable functions defined on $\mathcal{S}$ and therefore obey the closure relation
\begin{equation}
\sum_{n}\psi_{n}(\kappa,\boldsymbol{\rho})
\psi_{n}^{*}(\kappa,\boldsymbol{\rho}')
=\delta_{\mathcal{S}}(\boldsymbol{\rho}-\boldsymbol{\rho}'),
\label{op15}
\end{equation}
where $\delta_{\mathcal{S}}(\boldsymbol{\rho}-\boldsymbol{\rho}')$ is the Dirac delta function on $\mathcal{S}$.

Combining the definition of the DtN operator (\ref{op3}) with 
equation (\ref{op9}) we may write:
\begin{equation}
\hat{\mathcal{B}}(\kappa)\psi_{n}(\kappa,\boldsymbol{\rho})
=b_{n}(\kappa)\psi_{n}(\kappa,\boldsymbol{\rho}).
\label{op16}
\end{equation}
We observe that eigenvalues of the DtN operator are the eigenvalues 
$\left\{ b_{n}(\kappa)\right\}$ of the eigensystem  (\ref{op7})--(\ref{op9}) and the associated eigenfunctions are the surface parts $\left\{ \psi_{n}(\kappa,\boldsymbol{\rho})\right\}$ of the eigenfunctions $\left\{ \psi_{n}(\kappa,\mathbf{r})\right\}$. According to equation (\ref{op5}), we may rewrite equation (\ref{op16}) as
\begin{equation}
\int_{\mathcal{S}}\mathrm{d}\boldsymbol{\rho}' \, \mathcal{B}(\kappa,\boldsymbol{\rho},\boldsymbol{\rho}')\psi_{n}(\kappa,\boldsymbol{\rho}')
=b_{n}(\kappa)\psi_{n}(\kappa,\boldsymbol{\rho}).
\label{op17}
\end{equation}
Multiplying the above formula by $\psi_{n}^{*}(\kappa,\boldsymbol{\rho}'')$, summing over $n$ and using the closure relation 
(\ref{op15}) leads us to the spectral expansion of the DtN operator kernel
\begin{equation}
\mathcal{B}(\kappa,\boldsymbol{\rho},\boldsymbol{\rho}')
=\sum_{n}\psi_{n}(\kappa,\boldsymbol{\rho})
b_{n}(\kappa)\psi_{n}^{*}(\kappa,\boldsymbol{\rho}').
\label{op18}
\end{equation}
As the DtN operator and the NtD operator are mutually reciprocal, 
the spectral expansion of the NtD kernel takes form
\begin{equation}
\mathcal{R}(\kappa,\boldsymbol{\rho},\boldsymbol{\rho}')
=\sum_{n}\psi_{n}(\kappa,\boldsymbol{\rho})
b_{n}^{-1}(\kappa)\psi_{n}^{*}(\kappa,\boldsymbol{\rho}').
\label{op19}
\end{equation}
It is obvious from the expansions (\ref{op18}) and (\ref{op19}) that 
the operators $\hat{\mathcal{B}}(\kappa)$ and $\hat{\mathcal{R}}(\kappa)$ are Hermitian.
\section{The DtN method}

If the trial functions employed in the functional (\ref{eq30}) are continuous on $\mathcal{S}$, i.e.
\begin{equation}
\overline{\Psi}_{I}(\boldsymbol{\rho})
=\overline{\Psi}_{II}(\boldsymbol{\rho}),
\label{d1}
\end{equation}
the functional reduces to the following form:
\begin{equation}
 \mathcal{F}^{(D)}[\overline{\Psi}_{I},\overline{\Psi}_{II}]
=-\frac{\big<\overline{\Psi}_{I}\big|
\Delta\overline{\Psi}_{I}\big>_{I}
+\big<\overline{\Psi}_{II}\big|
\Delta\overline{\Psi}_{II}\big>_{II}}
{\big<\overline{\Psi}_{I}\big|\overline{\Psi}_{I}\big>_{I}
+\big<\overline{\Psi}_{II}\big|
\overline{\Psi}_{II}\big>_{II}} 
+\frac{\big(\overline{\Psi}_{I}\big|\nabla_{\perp}\overline{\Psi}_{I}
-\nabla_{\perp}\overline{\Psi}_{II}\big)}
{\big<\overline{\Psi}_{I}\big|\overline{\Psi}_{I}\big>_{I}
+\big<\overline{\Psi}_{II}\big|
\overline{\Psi}_{II}\big>_{II}}.
\label{d2}
\end{equation}
Let us assume that the trial function $\overline{\Psi}_{II}$ is some function $\psi^{(D)}(\kappa,\mathbf{r})$, obeying (\ref{op1}) and (\ref{op2}):
\begin{equation}
\overline{\Psi}_{II}(\mathbf{r})
=\psi^{(\mathrm{D})}(\kappa,\mathbf{r})
\qquad (\mathbf{r}\in\Gamma_{II}).
\label{d3}
\end{equation}
Such choice of $\overline{\Psi}_{II}(\mathbf{r})$ in virtue of equations (\ref{op3}) and (\ref{d1}) gives
\begin{equation}
\nabla_{\perp}\overline{\Psi}_{II}(\boldsymbol{\rho})
=\hat{\mathcal{B}}(\kappa)\overline{\Psi}_{I}(\boldsymbol{\rho}).
\label{d4}
\end{equation}
Applying equations (\ref{d3}), (\ref{op1}) and (\ref{d4}) to equation (\ref{d2}) leads us 
to such a form of the functional in which the only term containing 
$\psi^{(D)}(\kappa,\mathbf{r})$ is the integral $\big< \psi^{(D)} \big| \psi^{(D)} \big>_{II}$:
\begin{equation}
 \mathcal{F}^{(D)}[\overline{\Psi}_{I},\psi^{(D)}]
=-\frac{\big<\overline{\Psi}_{I}\big|
\Delta\overline{\Psi}_{I}\big>_{I}
-\kappa^{2}\big< \psi^{(D)} \big| \psi^{(D)} \big>_{II}}
{\big<\overline{\Psi}_{I}\big|\overline{\Psi}_{I}\big>_{I}
+\big< \psi^{(D)} \big| \psi^{(D)} \big>_{II}} 
+\frac{\big(\overline{\Psi}_{I}\big|\nabla_{\perp}\overline{\Psi}_{I}
-\hat{\mathcal{B}}\,\overline{\Psi}_{I}(\boldsymbol{\rho})\big)}
{\big<\overline{\Psi}_{I}\big|\overline{\Psi}_{I}\big>_{I}
+\big< \psi^{(D)} \big| \psi^{(D)} \big>_{II}}.
\label{d5}
\end{equation}
Subtracting the complex conjugation of the Helmholtz equation for $\psi^{(D)}(\kappa,\mathbf{r})$ multiplied by $\partial \psi^{(D)}(\kappa,\mathbf{r})/ \partial \kappa$ from the Helmholtz equation for $\psi^{(D)}(\kappa,\mathbf{r})$ differentiated with respect to $\kappa$ and multiplied by $\psi^{(D)*}(\kappa,\mathbf{r})$ we obtain
\begin{equation}
\psi^{(D)*}(\kappa,\mathbf{r}) \Delta 
\frac{\partial \psi^{(D)}(\kappa,\mathbf{r})}{\partial \kappa}
-\frac{\partial \psi^{(D)}(\kappa,\mathbf{r})}{\partial \kappa}
\Delta \psi^{(D)*}(\kappa,\mathbf{r})
=-2\kappa \psi^{(D)*}(\kappa,\mathbf{r})\psi^{(D)}(\kappa,\mathbf{r}).
\label{d6}
\end{equation}
Integration of (\ref{d6}) over $\Gamma_{II}$ after employing the Green's theorem, the definition (\ref{op3}), the Hermiticity of $\hat{\mathcal{B}}(\kappa)$ and the continuity constraint (\ref{d1}) results in
\begin{equation}
\big< \psi^{(D)} \big| \psi^{(D)} \big>_{II}= \frac{1}{2\kappa}
\bigg( \overline{\Psi}_{I} \bigg| \frac{\partial\hat{\mathcal{B}}}
{\partial\kappa} \overline{\Psi}_{I}\bigg).
\label{d7}
\end{equation}
Substitution of equation (\ref{d7}) transforms the functional (\ref{d5}) into the form
\begin{equation}
\hspace*{-1cm}\mathcal{F}^{(D)}[\kappa,\overline{\Psi}_{I}]
=\frac{-\big<\overline{\Psi}_{I}\big|
\Delta\overline{\Psi}_{I}\big>_{I}
+\big(\overline{\Psi}_{I} \big| \nabla_{\perp}\overline{\Psi}_{I}
-\hat{\mathcal{B}}\overline{\Psi}_{I}+\frac{\kappa}{2}
[{\partial\hat{\mathcal{B}}}/
{\partial\kappa}] \overline{\Psi}_{I}\big)}
{\big<\overline{\Psi}_{I}\big|\overline{\Psi}_{I}\big>_{I}
+\frac{1}{2\kappa}
\big( \overline{\Psi}_{I} \big| [{\partial\hat{\mathcal{B}}}/
{\partial\kappa}] \overline{\Psi}_{I}\big)},
\label{d8}
\end{equation}
which depends only on the parameter $\kappa$ and the trial function $\overline{\Psi}_{I}$ defined in $\Gamma_{I}$ and on $\mathcal{S}$. We arrive at a conclusion that the usage of the surface integral DtN operator allows us to reduce the initial problem defined in $\Gamma$ to the subdomain $\Gamma_{I}$ and the interface $\mathcal{S}$.

We need to find such functions $\Psi^{(D)}(\kappa,\mathbf{r})$, $(\mathbf{r} \in \Gamma_{I})$, that make the functional (\ref{d8}) stationary. 
The associated stationary values are estimates of some of the values $\left\{k^{2}\right\}$ 
appearing in the problem (\ref{eq1})--(\ref{eq2}). In practice, it may be impossible to find
$\Psi^{(D)}(\kappa,\mathbf{r})$ analytically. In order to obtain some approximate solutions let us represent the trial function $\overline{\Psi}_{I}(\mathbf{r})$ as a linear combination of some basis functions $\phi_{\mu}(\mathbf{r})$, defined in $\Gamma_{I}$:
\begin{equation}
\overline{\Psi}_{I}(\mathbf{r})
=\sum_{\mu=1}^{M}\overline{a}_{\mu}\phi_{\mu}(\mathbf{r})
\qquad (\mathbf{r}\in\Gamma_{I}).
\label{d9}
\end{equation}
Applying (\ref{d9}) to (\ref{d8}) yields
\begin{equation}
\mathcal{F}_{\phi}^{(D)}[\kappa,\overline{\mathsf{a}}\,^{\dag},
\overline{\mathsf{a}}]=\frac{\overline{\mathsf{a}}\,^{\dag}
\mathsf{\Lambda}^{(D)}(\kappa)\overline{\mathsf{a}}}
{\overline{\mathsf{a}}\,^{\dag}
\mathsf{\Delta}^{(D)}(\kappa)\overline{\mathsf{a}}},
\label{d10}
\end{equation}
where $\overline{\mathsf{a}}$ is an $M$-component column vector with elements $\{\overline{a}_{\mu} \}$ and $\overline{\mathsf{a}}\,^{\dag}$ is its Hermitian adjoint, while $\Lambda^{(D)}(\kappa)$ and $\Delta^{(D)}(\kappa)$ are $M\times M$ Hermitian matrices with elements 
\begin{equation}
\Lambda_{\mu\nu}^{(D)}(\kappa)
=\left[-\big<\phi_{\mu}\big|\Delta\phi_{\nu}\big>_{I}
+\big(\phi_{\mu}\big|\nabla_{\perp}\phi_{\nu}
-\hat{\mathcal{B}}\phi_{\nu}+\frac{\kappa}{2}
[\partial\hat{\mathcal{B}}/ \partial\kappa]\phi_{\nu}\big)\right],
\label{d11}
\end{equation}
\begin{equation}
\Delta_{\mu\nu}^{(D)}(\kappa)
=\big<\phi_{\mu}\big|\phi_{\nu}\big>_{I}
+\frac{1}{2\kappa}\big(\phi_{\mu}\big|
[\partial\hat{\mathcal{B}}/ \partial\kappa]\phi_{\nu}\big).
\label{d12}
\end{equation}
Let $\widetilde{\mathsf{a}}\,^{(D)}(\kappa)$ and $\widetilde{\mathsf{a}}\,^{(D)\dag}(\kappa)$ be such particular vectors $\overline{\mathsf{a}}$ and $\overline{\mathsf{a}}\,^{\dag}$ that make the functional (\ref{d10}) stationary with respect to variations in their components:
\begin{equation}
\delta \mathcal{F}_{\phi}^{(D)}[\kappa,\widetilde{\mathsf{a}}\,^{(D)\dag}(\kappa),
\widetilde{\mathsf{a}}\,^{(D)}(\kappa)]=0.
\label{d13}
\end{equation}
From equations (\ref{d13}) and (\ref{d10}) we arrive at the algebraic eigensystem
\begin{equation}
\mathsf{\Lambda}^{(D)}(\kappa)
\widetilde{\mathsf{a}}\,^{(D)}(\kappa)
=\widetilde{F}\,^{(D)}(\kappa)\mathsf{\Delta}^{(D)}(\kappa)
\widetilde{\mathsf{a}}\,^{(D)}(\kappa)
\label{d14}
\end{equation}
(and its Hermitian conjugate), where $\widetilde{F}\,^{(D)}(\kappa)$ is defined as
\begin{equation}
\widetilde{F}\,^{(D)}(\kappa)=\mathcal{F}_{\phi}^{(D)}
[\kappa,\widetilde{\mathsf{a}}\,^{(D)\dag}(\kappa),
\widetilde{\mathsf{a}}\,^{(D)}(\kappa)].
\label{d15}
\end{equation}
The eigensystem (\ref{d14}) has $M^{(D)}\leq M$ eigenvalues
$\{\widetilde{F}_{\gamma}\,^{(D)}(\kappa)\}$ and corresponding eigenvectors
$\{\widetilde{\mathsf{a}}_{\gamma}\,^{(D)}(\kappa)\}$. These eigenvalues are second-order variational estimates of some among values $\{k^{2}\}$ of the system (\ref{eq1})--(\ref{eq2}). In virtue of equation (\ref{d9}), the eigenvectors
$\{\widetilde{\mathsf{a}}_{\mu}\,^{(D)}(\kappa)\}$, 
with the components 
$\{\widetilde{a}_{\mu \gamma}^{\,(D)}(\kappa)\}$, give us $M^{(D)}$ functions:
\begin{equation}
\widetilde{\Psi}\,^{(D)}_{\gamma}(\kappa,\mathbf{r})
=\sum_{\mu=1}^{M}\widetilde{a}\,^{(D)}_{\mu\gamma}(\kappa)
\phi_{\mu}(\mathbf{r})
\qquad (\mathbf{r}\in\Gamma_{I}),
\label{d16}
\end{equation}
which are first-order variational estimates of some of the eigenfunctions of the system (\ref{eq1})--(\ref{eq2}) in the subdomain $\Gamma_{I}$. Now we may find the functions $\{\psi^{(D)}_{\gamma}(\kappa,\mathbf{r})\}$ which are the estimates of the eigenfunctions in $\Gamma_{II}$. Let us expand them in the basis constitued by the eigenfunctions of the Steklov system
(\ref{op7})--(\ref{op9}):
\begin{equation}
\psi^{(D)}_{\gamma}(\kappa,\mathbf{r})
=\sum_{n}c_{n\gamma}^{(D)}(\kappa)\psi_{n}(\kappa,\mathbf{r})
\qquad (\mathbf{r}\in\Gamma_{II}).
\label{d17}
\end{equation}
Letting the point $\mathbf{r}$ tend to the interface $\mathcal{S}$, employing the orthonormality relation (\ref{op14}), and using the formula (\ref{d1}), we obtain
\begin{equation}
c_{n\gamma}^{(D)}(\kappa)=\big(\psi_{n}\big|\widetilde{\Psi}\,^{(D)}_{\gamma}\big).
\label{d18}
\end{equation}
%
%
\section{The NtD method}
In the previous section we started our reasoning with the matching condition (\ref{d1}) for the trial functions used in the functional
(\ref{eq30}). Now, let us turn to another possibility and impose a weaker condition
\begin{equation}
\nabla_{\perp}\overline{\Psi}_{I}(\boldsymbol{\rho})
=\nabla_{\perp}\overline{\Psi}_{II}(\boldsymbol{\rho}).
\label{n1}
\end{equation}
In this case, the functional (\ref{eq30}) simplifies to 
\begin{equation}
\hspace*{-1cm}\mathcal{F}^{(N)}[\overline{\Psi}_{I},\overline{\Psi}_{II}]=-\frac{\big<\overline{\Psi}_{I}\big|
\Delta\overline{\Psi}_{I}\big>_{I}
+\big<\overline{\Psi}_{II}\big|
\Delta\overline{\Psi}_{II}\big>_{II}}
{\big<\overline{\Psi}_{I}\big|\overline{\Psi}_{I}\big>_{I}
+\big<\overline{\Psi}_{II}\big|
\overline{\Psi}_{II}\big>_{II}} 
-\frac{\big(\nabla_{\perp}\overline{\Psi}_{I}\big|
\overline{\Psi}_{I}-\overline{\Psi}_{II}\big)}
{\big<\overline{\Psi}_{I}\big|\overline{\Psi}_{I}\big>_{I}
+\big<\overline{\Psi}_{II}\big|
\overline{\Psi}_{II}\big>_{II}}.
\label{n2}
\end{equation}
We assume that the trial function $\overline{\Psi}_{II}(\mathbf{r})$ is some function obeying the set (\ref{op1})--(\ref{op2}):
\begin{equation}
\overline{\Psi}_{II}(\mathbf{r})
=\psi^{(N)}(\kappa,\mathbf{r})
\qquad (\mathbf{r}\in\Gamma_{II}).
\label{n3}
\end{equation}
Investigation analogous to that leading to (\ref{d7}) yields
\begin{equation}
\big< \psi^{(N)} \big| \psi^{(N)} \big>_{II}=-\frac{1}{2\kappa}
\bigg( \nabla_{\perp} \overline{\Psi}_{I} \bigg| 
\frac{\partial\hat{\mathcal{R}}}
{\partial\kappa} \nabla_{\perp} \overline{\Psi}_{I}\bigg)
\label{n4}
\end{equation}
and the functional (\ref{n2}) transforms to
\begin{equation}
\hspace*{-1cm}\mathcal{F}^{(N)}[\kappa,\overline{\Psi}_{I}]
=\frac{-\big<\overline{\Psi}_{I}\big|
\Delta\overline{\Psi}_{I}\big>_{I}
+\big(\nabla_{\perp}\overline{\Psi}_{I} \big| \hat{\mathcal{R}}\nabla_{\perp}\overline{\Psi}_{I}
-\overline{\Psi}_{I}-\frac{\kappa}{2}
[{\partial\hat{\mathcal{R}}}/{\partial\kappa}] \nabla_{\perp}\overline{\Psi}_{I}\big)}
{\big<\overline{\Psi}_{I}\big|\overline{\Psi}_{I}\big>_{I}
-\frac{1}{2\kappa}
\big( \nabla_{\perp}\overline{\Psi}_{I} \big| [{\partial\hat{\mathcal{R}}}/ {\partial\kappa}] \nabla_{\perp} \overline{\Psi}_{I}\big)} .
\label{n5}
\end{equation}
Following the method of algebraization applied in case of the DtN method (see the formulas
(\ref{d9})--(\ref{d14})) we arrive at the generalized matrix eigensystem
\begin{equation}
\mathsf{\Lambda}^{(N)}(\kappa)
\widetilde{\mathsf{a}}\,^{(N)}(\kappa)
=\widetilde{F}\,^{(N)}(\kappa)\mathsf{\Delta}^{(N)}(\kappa)
\widetilde{\mathsf{a}}\,^{(N)}(\kappa)
\label{n6}
\end{equation}
(and its Hermitian matrix conjugate), where $\mathsf{\Lambda}^{(N)}(\kappa)$ and
$\mathsf{\Delta}^{(N)}(\kappa)$ are $M \times M$ matrices with elements
\begin{equation}
\Lambda_{\mu\nu}^{(N)}(\kappa)
=\left[-\big<\phi_{\mu}\big|\Delta\phi_{\nu}\big>_{I}
+\big(\nabla_{\perp}\phi_{\mu}\big|\hat{\mathcal{R}}\nabla_{\perp}\phi_{\nu}
-\phi_{\nu}-\frac{\kappa}{2}
[\partial\hat{\mathcal{R}}/ \partial\kappa]\nabla_{\perp}\phi_{\nu}\big)\right]
\label{n7}
\end{equation}
and
\begin{equation}
\Delta_{\mu\nu}^{(N)}(\kappa)
=\big<\phi_{\mu}\big|\phi_{\nu}\big>_{I}
-\frac{1}{2\kappa}\big(\nabla_{\perp}\phi_{\mu}\big|
[\partial\hat{\mathcal{R}}/ \partial\kappa]\nabla_{\perp}\phi_{\nu}\big).
\label{n8}
\end{equation}
Eigenvalues $\{\widetilde{F}_{\gamma}\,^{(N)}(\kappa)\}$ of (\ref{n6}) are second-order variational estimates of some of the values $\{k^{2}\}$ appearing in the set (\ref{eq1})--(\ref{eq2}), while 
components of the associated eigenvectors
$\{\widetilde{\mathsf{a}}_{\gamma}\,^{(N)}(\kappa)\}$ yield the estimates of eigenfunctions of (\ref{eq1})--(\ref{eq2}) in $\Gamma_{I}$:
\begin{equation}
\widetilde{\Psi}\,^{(N)}_{\gamma}(\kappa,\mathbf{r})
=\sum_{\mu=1}^{M}\widetilde{a}\,^{(N)}_{\mu\gamma}(\kappa)
\phi_{\mu}(\mathbf{r}) \qquad (\mathbf{r}\in \Gamma_{I}).
\label{n9}
\end{equation}
The last step is to find estimates of eigenfunctions in $\Gamma_{II}$. We expand $\psi^{(N)}_{\gamma}(\kappa,\mathbf{r})$ as follows
\begin{equation}
\psi^{(N)}_{\gamma}(\kappa,\mathbf{r})
=\sum_{n}c_{n\gamma}^{(N)}(\kappa)\psi_{n}(\kappa,\mathbf{r})
\qquad (\mathbf{r}\in\Gamma_{II}).
\label{n10}
\end{equation}
The orthonormality relation (\ref{op14}), the properties of the 
NtD operator and the matching condition (\ref{n1}) lead to
\begin{equation}
c_{n\gamma}^{(N)}(\kappa)=b^{-1}_{n}(\kappa)\big(\psi_{n}\big|\nabla_{\perp}\widetilde{\Psi}^{(N)}_{\gamma}\big).
\label{n11}
\end{equation}

It is worth noticing that in general the DtN method and the NtD method will give different estimates of the solutions of the initial system. 

More details about the DtN and NtD methods (for bound states of the Schr\"odinger equation and the Dirac equation in $\mathbb{R}^{3}$) may be found in the works of Szmytkowski and Bielski \cite{Szmy04, Biel06}.

\section{Numerical example}

\begin{figure}[h]
\begin{center}
\includegraphics[width=5cm]{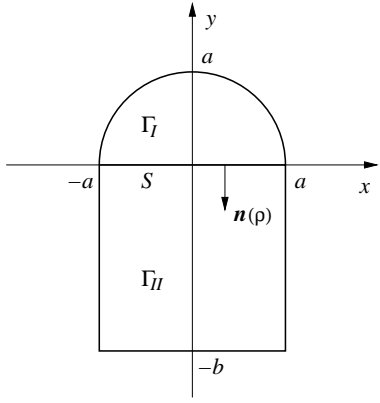}
\caption{\label{fig2}Geometry of the system used in the numerical
illustration.}
\end{center}
\end{figure}

To test the two methods, a few series of numerical calculations have been performed. A system in which $\Gamma$ is a two-dimensional domain consisting of a semicircle of radius $a$ joined to a rectangle of sides $a$ and $b$, as depicted in figure~\ref{fig2}, has been examined. 
The first step  is to decide which part of the whole domain $\Gamma$ should be $\Gamma_{I}$ and which one should be $\Gamma_{II}$. The decision depends on the simplicity of the construction of the DtN and the NtD operators. The region in which it is easier to solve (\ref{op7})--(\ref{op9}) should be taken as $\Gamma_{II}$. In our example the subdomain $\Gamma_{I}$ is the semicircle and the subdomain $\Gamma_{II}$ is the rectangle. It is not difficult to verify that the eigenvalues of (\ref{op7})--(\ref{op9}) in this case are
\begin{equation}
 b_{n}(\kappa)=  
\left\{ 
\begin{array} {ccc}
-\sqrt{\kappa^{2}-n^{2}\pi^{2}/4a^{2}}\,\cot(\sqrt{\kappa^{2}-n^{2}\pi^{2}/4a^{2}}\,b)
& \textrm{for} & \kappa^{2} \geq n^{2}\pi^{2}/4a^{2} \\
-\sqrt{n^{2}\pi^{2}/4a^{2}-\kappa^{2}}\,\coth(\sqrt{n^{2}\pi^{2}/4a^{2}-\kappa^{2}}\,b)
& \textrm{for} & k^{2} < n^{2}\pi^{2}/4a^{2} 
\end{array}
\right. , 
\label{num1}
\end{equation}
where $n=1,2,\ldots$. The corresponding eigenfunctions are of the form
\begin{eqnarray}
&&\psi_{n}(\kappa,x,y)=  \nonumber \\  &&
\quad A_{n} \sin\left[ \frac{n\pi(x+a)}{2a}\right]\left\{ 
\begin{array} {ccc}
\sin\left[\sqrt{\kappa^{2}-n^{2}\pi^{2}/4a^{2}}\,(y+b)\right]
& \textrm{for} & \kappa^{2} \geq n^{2}\pi^{2}/4a^{2} \\
\sinh\left[\sqrt{n^{2}\pi^{2}/4a^{2}-\kappa^{2}}\,(y+b)\right]
& \textrm{for} & \kappa^{2} < n^{2}\pi^{2}/4a^{2} 
\end{array}
\right. ,  \label{num2}
\end{eqnarray}
where $A_{n}$ according to the relation (\ref{op14}) are
\begin{equation}
\label{num3}
 A_{n}=\left\{ 
\begin{array} {ccc}
\left(\sqrt{a}\,\sin\left[\sqrt{\kappa^{2}-n^{2}\pi^{2}/4a^{2}}\,b\right]\right)^{-1}
& \textrm{for} & \kappa^{2} \geq n^{2}\pi^{2}/4a^{2} \\
\left(\sqrt{a}\,\sinh\left[\sqrt{n^{2}\pi^{2}/4a^{2}-\kappa^{2}}\,b\right]\right)^{-1}
& \textrm{for} & \kappa^{2} < n^{2}\pi^{2}/4a^{2} 
\end{array}
\right. . 
\end{equation}
Using equations (\ref{num1})--(\ref{num3}) in (\ref{op18}) and (\ref{op19}), we obtain the kernels of the DtN and NtD operators. Let us observe that in the examined case we may distinguish the even (symmetric with respect to the y-axis) and the odd (antisymmetric) modes. We may search for them separately (which means working with smaller matrices), applying apropriate basis functions $\{\phi_{\mu}(\mathbf{r})\}$ in
(\ref{d9}). For the even modes we may use
\begin{eqnarray}
\label{num4}
&&\phi_{1}(\mathbf{r})=r-a, \quad \mathbf{r} \in \Gamma_{I},\\
\label{num5}
&&\phi_{\mu}(\mathbf{r})=\phi_{nm}(\mathbf{r})=r\sin[n\alpha(r-a)]\cos(m\beta\varphi), \quad \mathbf{r} \in \Gamma_{I} \\ &&\qquad \qquad \qquad(\mu=2,3,\ldots, \quad n,m=1,2,\ldots) \nonumber
\end{eqnarray}
and for the odd modes we may apply
\begin{eqnarray}
\label{num6}
&&\phi_{\mu}(\mathbf{r})=\phi_{nm}(\mathbf{r})=r\sin[n\alpha(r-a)]\sin(m\beta\varphi), \quad \mathbf{r} \in \Gamma_{I} \\ && \qquad \qquad \qquad (\mu=1,2,\ldots, \quad n,m=1,2,\ldots) \nonumber
\end{eqnarray}
(each $\mu$ represents an unique combination of two integers $n$ and $m$). In the above formulas $\varphi$ is the angle between the y-axis and the position vector $\mathbf{r}=[x,y]$ (we assume that $\varphi$ is positive for $x<0$ and negative for $x>0$), $r$ is the length of $\mathbf{r}$, while $\alpha$ and $\beta$ are some arbitrary real parameters. Note that all the functions vanish on $\partial \Gamma_{I}$. The variational bases are formed from the functions with $1 \leq n \leq n_{max}$ and $1 \leq m \leq m_{max}$. Do not forget about the extra function (\ref{num4}) used for the symmetric states. The function is added to the basis because all the functions (\ref{num5}) are equal to zero at $r=0$, while in general the eigenfunctions of the even modes may be nonzero at $r=0$.

To obtain the estimates of some $k^{2}$ and $\Psi(\mathbf{r})$, we must establish an initial value of $\kappa$ (which is some estimate of $k$). 
Then we apply some chosen variational basis, calculate the matrix elements of
$\mathsf{\Lambda}^{(D)}(\kappa)$ and $\mathsf{\Delta}^{(D)}(\kappa)$ or
$\mathsf{\Lambda}^{(N)}(\kappa)$ and $\mathsf{\Delta}^{(N)}(\kappa)$ and solve the matrix system
(\ref{d14}) or (\ref{n6}). The resulting eigenvalues $\{\widetilde{F}_{\gamma}\,^{(D \; or \;
N)}(\kappa)\}$ are used to set new values of $\kappa$ (separately for each $\gamma$). We focus ourselves on an arbitrary chosen state, let it be the state with $\gamma=\gamma'$, so we take 
\begin{equation}
\label{num7}
\kappa=\sqrt{\widetilde{F}_{\gamma'}\,^{(D \; or \; N)}(\kappa)}.
\end{equation}
We find new matrices, solve the new matrix system and obtain new estimates of eigenvalues $\{\widetilde{F}_{\gamma}\,^{(D \; or \;
N)}(\kappa)\}$ with $\widetilde{F}_{\gamma'}\,^{(D \; or \;
N)}(\kappa)$ among them. We then apply (\ref{num7}) again and the iterative procedure repeats until convergence of
$\widetilde{F}_{\gamma'}\,^{(D \; or \;
N)}(\kappa)$ is achieved. 

\begin{figure}[h]
\begin{center}
\begin{tabular}{cc}
\includegraphics[width=5.5cm,angle=-90]{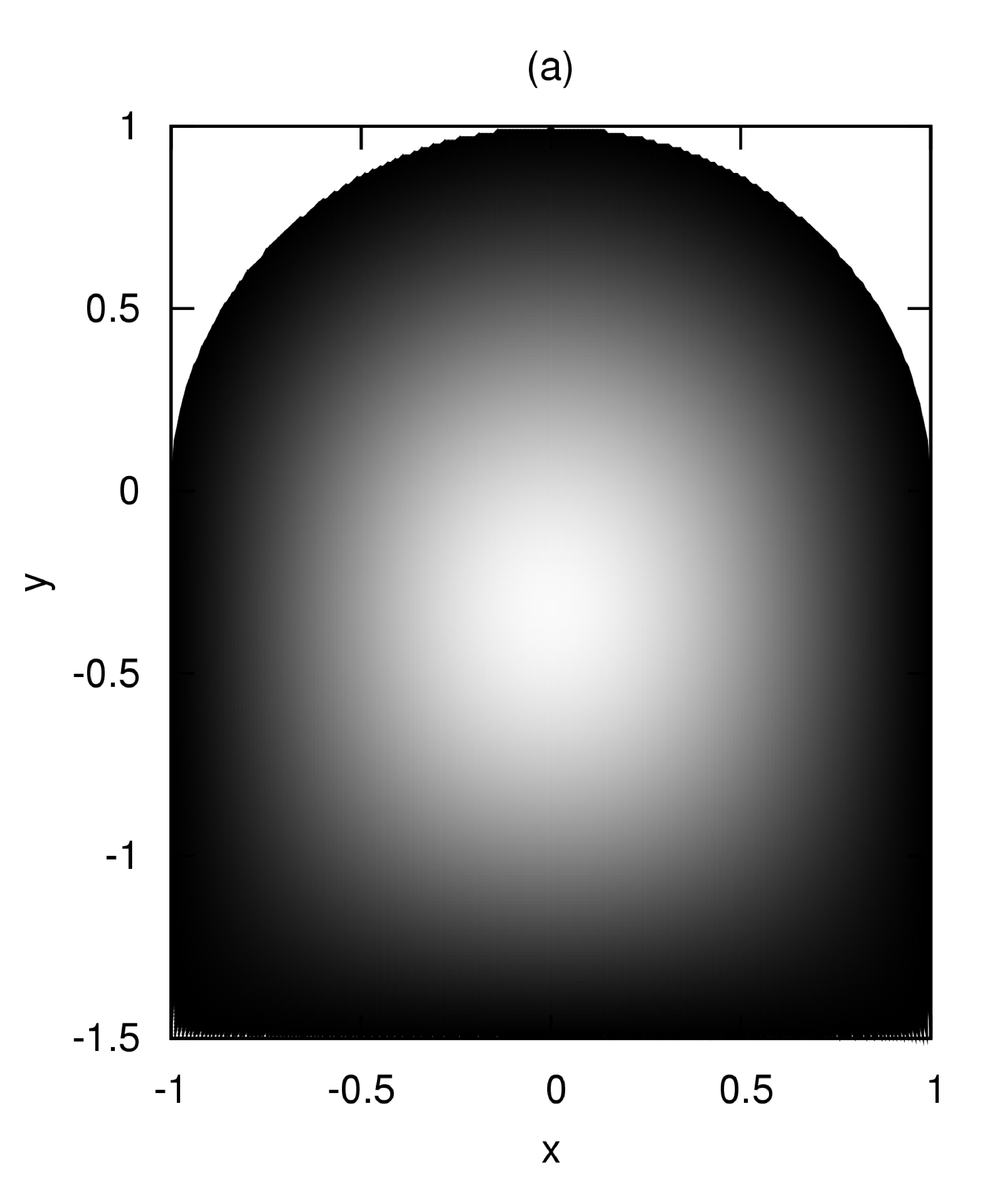} &
\includegraphics[width=5.5cm,angle=-90]{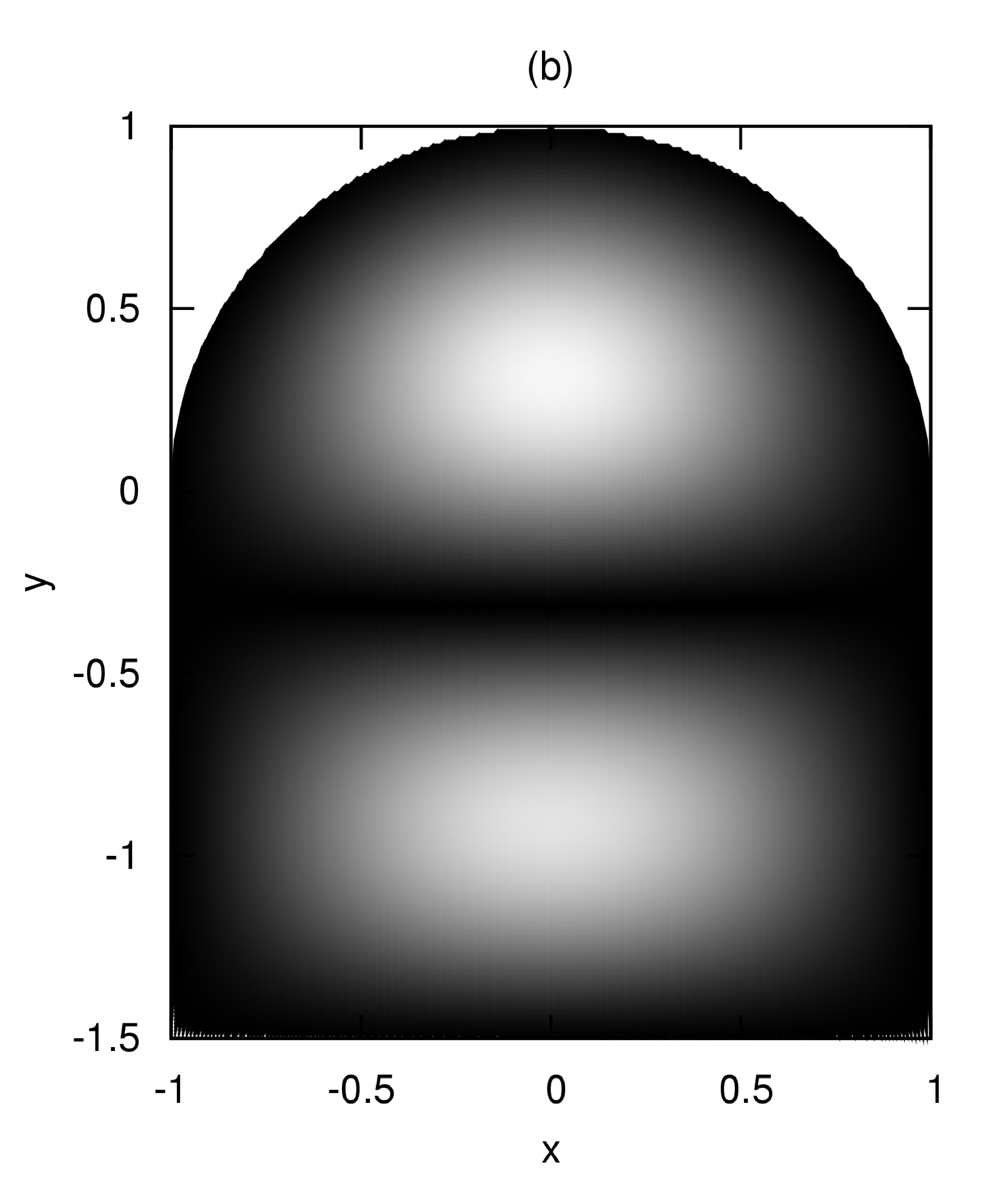}\\
\includegraphics[width=5.5cm,angle=-90]{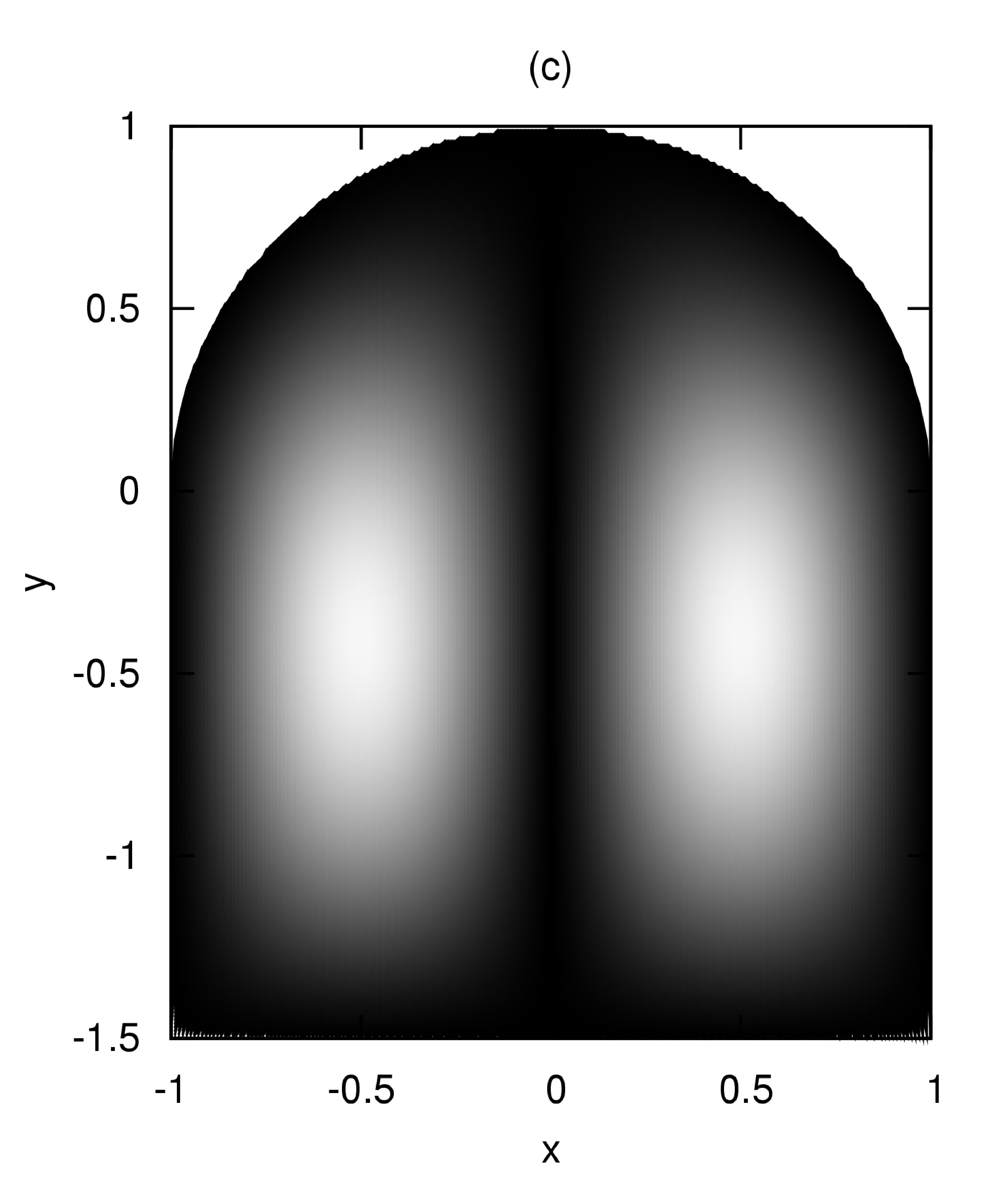} &
\includegraphics[width=5.5cm,angle=-90]{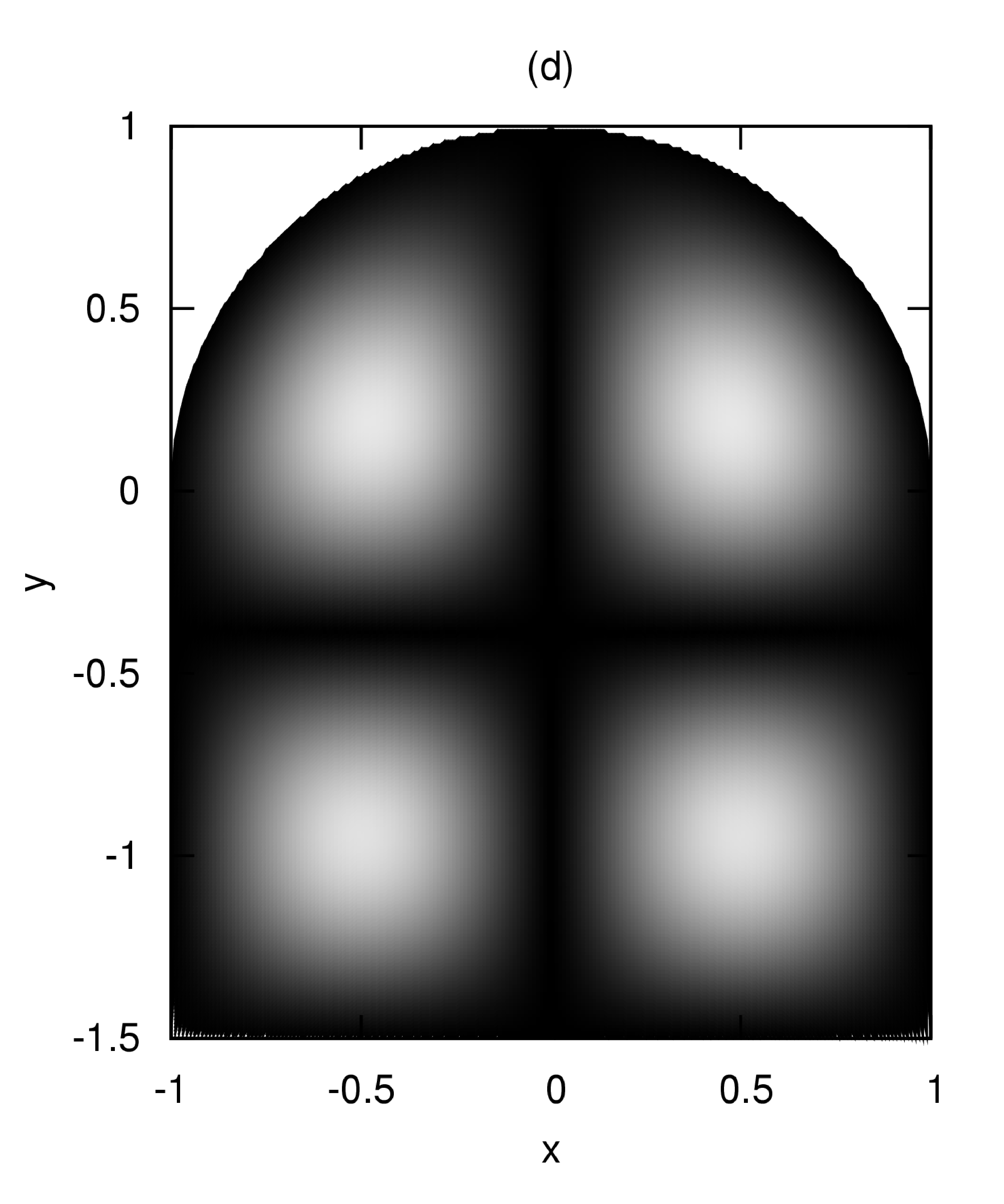}
\end{tabular}
\caption{\label{fig3} Density plots of $|\Psi|^{2}$: (a) and (b) -- the two lowest even modes; (c) and (d) -- the two lowest odd modes. The results obtained by the DtN method (see section IV). The basis functions (\ref{num4})--(\ref{num6}) with $1\leq n \leq 25$ and $1\leq m \leq 25$ have been used. The converged estimates of $k$ (see the last row of table~\ref{tab2}) have been taken as the values of $\kappa$. }
\end{center}
\end{figure}

\begin{table}[h]
\begin{center}
\begin{tabular}{|c|c|c|c|c|}
\hline
Iteration & 
$k_{even,1}^{(D)}$ & $k_{even,1}^{(N)}$ & $k_{odd,1}^{(D)}$ & $k_{odd,1}^{(N)}$ \\
\hline
1 & 2.0633 & 2.0487 & 3.4586 & 3.4200 \\
\hline
2 & 2.0611 & 2.0604 & 3.4508 & 3.4447 \\
\hline
3 & 2.0611 & 2.0611 & 3.4507 & 3.4505 \\
\hline
4 &  & 2.0611 & 3.4507 & 3.4507 \\
\hline
5 &  &  &  & 3.4507 \\
\hline
\end{tabular}
\caption{\label{tab1}Convergence rate of the DtN and the NtD variational estimates of $k$ of the lowest even mode and the lowest odd mode of the system used in the numerical example. The results obtained by employing the basis functions (\ref{num4})--(\ref{num6}) with $1 \leq n \leq 15$ and $1 \leq m \leq 15$. The inputs for the iteration procedure have been $\kappa=2.0116$ for the first even mode and $\kappa=3.3836$ for the first odd mode. SI units are used.}
\end{center}
\end{table}
\begin{table}[h]
\begin{center}
\begin{tabular}{|c|c|c|c|c|}
\hline
$n_{max}$,$m_{max}$ &$k_{even,1}^{(D)}$ & $k_{even,2}^{(D)}$&$k_{odd,1}^{(D)}$ &$k_{odd,2}^{(D)}$\\
 & $k_{even,1}^{(N)}$ & $k_{even,2}^{(N)}$&$k_{odd,1}^{(N)}$ &$k_{odd,2}^{(N)}$\\
\hline
3,3  & 2.0630 & 3.0745 & 3.4527 & 4.2234 \\
     & 2.0628 & 3.0809 & 3.4527 & 4.2234 \\
\hline
5,5  & 2.0611 & 3.0734 & 3.4511 & 4.2200 \\
     & 2.0611 & 3.0734 & 3.4511 & 4.2200 \\
\hline
15,15& 2.0611 & 3.0731 & 3.4507 & 4.2190 \\
     & 2.0611 & 3.0731 & 3.4507 & 4.2190 \\
\hline
25,25& 2.0611 & 3.0730 & 3.4506 & 4.2189 \\
     & 2.0611 & 3.0730 & 3.4506 & 4.2189 \\
\hline
30,30& 2.0611 & 3.0730 & 3.4506 & 4.2189 \\
     & 2.0611 & 3.0730 & 3.4506 & 4.2189 \\
\hline
\end{tabular}
\caption{\label{tab2}Converged DtN and NtD variational estimates of $k$ of the two lowest even modes and the two lowest odd modes. The basis functions (\ref{num4})--(\ref{num6}) with $1\leq n \leq n_{max}$ and $1\leq m \leq m_{max}$ have been used. The inputs for the iterative procedures have been $\kappa=2.0116$ and $\kappa=2.9638$ for the even modes and $\kappa=3.3836$ and $\kappa=4.0232$ for the odd modes. SI units are used.}
\end{center}
\end{table}
The numerical calculations have been made for $a=1$, $b=1.5$ and $\alpha=\beta=1$ (SI units are used).
Table~\ref{tab1} presents estimates of $k$ of the first even mode and the first odd mode obtained by using the iterative procedure described above. The variational bases have contained the functions (\ref{num4}) and (\ref{num5}) (for the even mode) or (\ref{num6}) (for the odd mode) with $1\leq n \leq 15$ and $1\leq m \leq 15$. The initial values of $\kappa$ have been $\kappa=2.0116$ for the first even mode and $\kappa=3.3836$ for the first odd mode, they are the exact values of $k$ (with accuracy of 4 decimal places) of the first even mode and the first odd mode of the system (\ref{eq1})--(\ref{eq2}) with $\Gamma$ being a rectangle of sides 2 and 2.5 (in the next series of calculations the fact that in case of the rectangle the second even mode is characterized by $k=2.9638$ and the second odd mode by $k=4.0232$ has been used). We see that the estimates converge after a few iterations, and that the results obtained by the DtN method converge faster. To verify, how the converged (after the iterative procedure) estimates of $k$ depend on the basis dimensions, let us analyze the results collected in table~\ref{tab2}. The two lowest even modes and the two lowest odd modes of our problem have been examined. The inputs for the iterative procedures have been $\kappa=2.0116$ and $\kappa=2.9638$ for the even modes and $\kappa=3.3836$ and $\kappa=4.0232$ for the odd modes. It is seen that even quite small bases lead to estimates of good accuracy.
The aim of the next series of numerical calculations was to estimate eigenfunctions $\Psi$ of the four states.
Figure~\ref{fig3} shows density plots of $|\Psi|^{2}$ of the four modes obtained by using the DtN method in a way presented at the end of section IV. In the domain $\Gamma_{I}$ the basis functions (\ref{num4})--(\ref{num6}) with $1\leq n \leq 25$ and $1\leq m \leq 25$ have been used. The values of $\kappa$ have been equal to the converged estimates of $k$ (see table~\ref{tab2}). Density plots obtained by the NtD method are exactly the same.
\\*[1cm]
\textbf{\large Acknowledgments}
\\*[1cm]
I wish to thank Professor R.~Szmytkowski for many useful discussions.
\newpage

\begin{thebibliography}{20}
\bibitem{Amor08}
P. Amore, J. Phys. A: Math. Theor. {\bf 41}, 265206/1--29 (2008)
%
\bibitem{Chak09}
S. Chakraborty, J. K. Bhattacharjee, S. P. Khastgir, J. Phys. A: Math. Theor. {\bf 42}, 195301/1--12 (2009)
%
\bibitem{Stei10}
O. Steinbach, M. Windisch, Numer. Math. DOI: 10.1007/s00211-010-0315-6 (2010)
%
\bibitem{Szmy04} 
R. Szmytkowski, S. Bielski, Phys. Rev. A {\bf 70}, 042103/1--12 (2004)
%
\bibitem{Biel06}
S. Bielski, R. Szmytkowski,  J. Phys. A {\bf 39}, 7359--7381 (2006)
%
\bibitem{Ingl81}
J. E. Inglesfield, J. Phys. C {\bf 14}, 3795--3806 (1981)
%
\bibitem{Szmy97}
R. Szmytkowski, J. Phys. A: Math. Gen. {\bf 30}, 4413--4438 (1997)
%
\bibitem{Szmy98}
R. Szmytkowski, J. Math. Phys. {\bf 39}, (1998) 5231--5252, Erratum: J. Math. Phys. {\bf 40}, 4181 (1999)
%
\bibitem{Gerj83}
E. Gerjuoy, A. R. P. Rau, L. Spruch, Rev. Mod. Phys. {\bf 55}, 725--774 (1983)
%
\bibitem{Szmy04a}
R. Szmytkowski, S. Bielski, Int. J. Quantum Chem. {\bf 97}, 966--976 (2004)
%
\bibitem{Auch04}
G.\ Auchmuty, 
Num. Funct. Anal. Opt. {\bf 25}, 321--348 (2004)
\end{thebibliography}
\end{document}